\documentclass{acm_proc_article-sp}

\usepackage[usenames,dvipsnames]{color}
\usepackage{amssymb}
\usepackage{multirow}
\usepackage{hyperref}
\usepackage{tikz}
\usepackage{subcaption}
\usepackage{graphicx}

\newcommand{\secref}[1]{Section~\ref{sec_#1}}
\newcommand{\figref}[1]{Figure~\ref{fig_#1}}

\definecolor{red}{RGB}{147, 82, 0}
\definecolor{green}{RGB}{146, 144, 0}
\definecolor{blue}{RGB}{0, 84, 147}
\definecolor{gray}{RGB}{121, 121, 121}

\makeatletter
\def\@copyrightspace{\relax}
\makeatother

\begin{document}


\title{Package equivalence in complex software network}

\numberofauthors{2}
\author{
\alignauthor
Tomislav Slijep{\v c}evi\'c\\
       \affaddr{University of Ljubljana}\\
	\affaddr{Faculty of Computer and Information Science}\\
       \affaddr{Ve\v{c}na pot 113, SI-1000 Ljubljana, Slovenia}\\
       \email{ts2287@student.uni-lj.si}
}

\date{\today}
\maketitle


\begin{abstract} 

The public package registry \textit{npm} is one of the biggest software registry. With its 216 911 software packages, it forms a big network of software dependencies. In this paper we evaluate various methods for finding similar packages in the \textit{npm} network, using only the structure of the graph. Namely, we want to find a way of categorizing similar packages, which would be useful for recommendation systems. This size enables us to compute meaningful results, as it softened the particularities of the graph. \textit{Npm} is also quite famous as it is the default package repository of \textit{Node.js}. We believe that it will make our results interesting for more people than a less used package repository. This makes it a good subject of analysis of software networks.

\end{abstract}






\section{Introduction}
\label{sec_intro}
The major interest of the network science is the analysis of the structural and statistical properties of complex networks. By looking at the patterns of connectivity in a network, a "role" can be assigned to each node~\cite{borgatti1992notions,wasserman1994social}. Researches have discovered that there is a close relation between the structure of a network and the roles, while they analyzed various networks that regards the life science, ecology, information, social sciences and economics~\cite{newman2003structure, barabasi2004network, jeong2000large, jeong2001lethality}. That is why it is important to understand the structure, i.e., the topology, of a network to find the roles and to understand the dynamics of a network in that matter~\cite{reichardt2007role}.

The node gets a role based on how it interacts with others. For the role assignment there have been developed two basic concepts: structural~\cite{LW71} and regular equivalence~\cite{WR83}. When two nodes are either structurally or regularly equivalent, then they have the same role in a network. For two nodes to be structurally equivalent, they have to share the same neighbors. The regular equivalence is a looser criterion than the structural equivalence; it does not compare the neighborhoods of the two nodes, but whether the two nodes are connected in the same way to the others. It holds that if two nodes are structurally equivalent, then they are also regularly equivalent, but not vice versa. The grouping of nodes that are related by either of equivalence relations forms equivalence classes, which represent the roles of nodes in a network.

In this paper we examine the complex software network to see if nodes can be naturally grouped into equivalence classes just as in other networks. The analyzed network consists of nodes that represent software packages and edges that indicate dependencies among packages. The data was taken from the biggest public software registry \textit{npm}~\cite{modulecount} (~\url{https://npmjs.com}). We believe that applying the relation of structural equivalency on the software network would group the software packages into equivalence classes that represent different types of software, e.g. a class of packages that is related to web, graphics or other. Reasoning behind this is that packages that are related by type share many neighbors and would therefore be assigned to the same equivalence class. By applying the other equivalence relation, which is the regular equivalence, we assume that the packages will be grouped into the following three equivalence classes: a class of core libraries, a class of the popular frameworks, and a class of packages that are supplementary to the popular frameworks, i.e. packages that serve as an add-on to one of the popular framework. Packages within listed classes connects to others in the same way and that is why we predict the formation of these classes. For a class of the core libraries it is understandable that they do not have dependencies, as there is not any package to depend upon at start, but they do have a lot of dependents. For the packages within the second class, the popular frameworks, it reasonable that they have a lot of dependencies and a lot of dependents. The packages in the last class, which supplement the popular frameworks, should have a few dependencies and probably none dependents.

Thus it should be possible to find similar packages using only the structure of the network. This could lead to amelioration of package recommendations based on equivalence relations and also to automatic labeling/categorizing of a new package based just on the listed dependencies. This system could help to homogenize keywords in software registry and thus improve the referencing of the packages and simplify the search for packages.

The rest of the article is structured as follows. In~\secref{rel_work} we present related work and in~\secref{meth} we give a formal introduction to node equivalence. Empirical evaluation with discussion is done in~\secref{res} and conclusions in~\secref{conc}.


\section{Related work}
\label{sec_rel_work}
The modern science of network is particularly interested in decomposing nodes of large networks into independent groups called "communities"~\cite{girvan2002community}. As a community, you think of a group of nodes that is internally densely connected but sparsely connected externally~\cite{reichardt2007role,wasserman1994social,girvan2002community,newman2003structure}. The nodes within community are similar to each other and dissimilar to the rest of the nodes in a network. Researchers showed there is actually community structures in real-world networks~\cite{flake2000efficient,girvan2002community,guimera2005functional}. However, the definition of community is not universally accepted and for that matter we have multiple definitions. Community structure emphasizes cohesive groups of nodes and the absence of dependencies between the groups, but this does not say anything about the roles in a network. The concept of roles in networks is much wider that the concept of community. Prerequisite for the analysis of roles in a network is a community structure. After that, you examine how discovered communities are inter-dependent, which translates to different roles in a network.

One of predominant technique for deriving structure of a network is blockmodeling~\cite{LW71}. It is considered as mapping nodes and edges onto their images in a reduced graph. A node ("block") in reduced graph represents the nodes from original graph that were mapped to the same image. This type of mapping is called semigroup homomorphism~\cite{WR83}. Homomorphisms are mappings that preserve the structure of a graph in a way that nodes of a graph are mapped into nodes in an image of the graph, and each edge in an image of the graph is connected to the same nodes as in the original graph. The semigroup is an algebraic structure consisting of a set with an associative binary operation. We can look at the nodes from original graph as a semigroup. Consequently, the nodes in reduced graph represent the result of applying the semigroup operation to the ordered pair $(x,y)$, which in our case are the edges in the original graph. If we take equivalence relation as the operation within semigroup, then we get equivalence classes as the nodes in the reduced graph. The plot of adjacency matrix of the new graph reveals blocks in diagonal, which are equivalence classes, and dependencies between them.

There exists a framework for blockmodeling classes within complex networks~\cite{reichardt2007role}. Authors derive a measure to find the best fitting image graph (reduced graph) and present a criterion to avoid overfitting. Image graph is termed as the role model. They did not demand exact mapping of every single node to the role model, but that the network as a whole fit as well as possible to the role model. The perfect fit would correspond to regular and structural equivalence. When tackling a new network, they assumed a given image graph and assignment of roles to nodes. They derived a quality function as an objective measure of fit between the image and the network under this assignment of roles. The assignment of roles which maximizes their quality function is considered as the best one to describe the connection structure of the original network. The concepts of modularity~\cite{newman2004finding} and structural equivalence are found as special cases of this approach. The modularity measure is commonly used in community detection algorithms. It measures the quality of a network division into communities.
The proposed method is applicable to both two-mode and one-mode data, directed and undirected, as well as weighted networks. It is non-parametric and computationally efficient. In the same paper authors applied method to the world trade network and analyzed the roles individual countries play in the global economy.

Another research dealt with analysis of patterns of role-to-role connections, but with their own definition of roles~\cite{guimera2007}. Principle was the same; to group the nodes into roles, according to their pattern of intra- and intergroup connections. They analyzed four different types of real-world networks; metabolic networks, protein interactions, global and regional air transportation networks, and the Internet at the level of autonomous systems. To determine and quantify the modular structure of these networks, they use simulated annealing. The optimal partitions of the network into groups was found with the use of the modularity measure. By comparing modular structure of each network with the randomization of the same network, they found out that all observed networks have a significant modular structure. That is reasonable as groups in biological networks corresponds to functional units and in air transportation groups corresponds to geo-political units. The role of each node was determined according to two properties: the relative within-group degree $z$, which measures the node's degree of connectedness with the nodes within the same group, and the participation coefficient $P$, which measures node's degree of connectedness with other groups. When classifying nodes into roles, they initially measured within-group degree of all nodes, and divided them into hubs and non-hubs based on high and low within-module degree, respectively. Then they looked at the participation coefficient to further subdivide hubs into different types of hubs. Further subdivision was also done to non-hubs. Reasoning behind this particular definition of the roles is given in~\cite{guimera2005functional}.

\section{Methods}
\label{sec_meth}

\begin{figure*}
\centering
\begin{minipage}{.3\textwidth}
\centering
\begin{tikzpicture}
  [scale=.8,auto=left,every node/.style={circle,fill=blue!20}]
  \node (a) at (4,5)  {a};
  \node (b) at (2,3)  {b};
  \node (c) at (5,3)  {c};
  \node [fill=red!20] (d) at (1,1)  {d};
  \node [fill=red!20] (e) at (3,1)  {e};
  \node (f) at (5,1)  {f};

  \path[every node/.style={font=\sffamily\small}]
	(a) edge node {} (b)
        edge node {} (c)
	(b) edge node {} (d)
        edge node {} (e)
	(c) edge node {} (f);
\end{tikzpicture}
\caption{Structurally equivalent nodes are marked red.}
\label{fig_structural_eq}
\end{minipage}%
\hspace*{0.3cm}
\begin{minipage}{.3\textwidth}
\centering 
\begin{tikzpicture}
  [scale=.8,auto=left,every node/.style={circle,fill=blue!20}]
  \node [fill=red!20] (a) at (4,5)  {a};
  \node [fill=yellow!20] (b) at (2,3)  {b};
  \node [fill=yellow!20] (c) at (5,3)  {c};
  \node (d) at (1,1)  {d};
  \node (e) at (3,1)  {e};
  \node (f) at (5,1)  {f};

  \path[every node/.style={font=\sffamily\small}]
	(a) edge node {} (b)
        edge node {} (c)
	(b) edge node {} (d)
        edge node {} (e)
	(c) edge node {} (f);
\end{tikzpicture}
\caption{Regularly equivalent nodes are those sharing the same color.}
\label{fig_regular_eq}
\end{minipage}%
\hspace*{0.3cm}
\begin{minipage}{.3\textwidth}
\centering 
\begin{tikzpicture}
  [scale=.8,auto=left,every node/.style={circle,fill=blue!20}]
  \node [fill=yellow!20] (a) at (3,5)  {a};
  \node (b) at (2,1)  {b};
  \node (c) at (4,1)  {c};
  \node (n1) at (1,3)  {};
  \node (n2) at (3,3)  {};
  \node (n3) at (5,3)  {};

  \path[every node/.style={font=\sffamily\small}]
	(a) edge node {} (n2)
        edge node {} (n3)
	(b) edge node {} (n1)
        edge node {} (n2)
        edge node {} (n3)
	(c) edge node {} (n3);
\end{tikzpicture}
\caption{Problem of basic similarity measure for structural equivalence which just counts the number of common neighbors.}
\label{fig_structural_eq_basic_measure_prob}
\end{minipage}
\end{figure*}

A graph is an ordered pair $G = \langle P, R \rangle$, where $P$ is a finite set of points and $R$ is a relation on $P$, i.e., a subset of the ordered pairs of points in $P \times P$. An equivalence $\equiv$ on $P$ is a relation such that for all $a, b, c \in P$, it has a property of reflexivity ($a \equiv a$), symmetry ($a \equiv b \implies b \equiv a$) and transitivity ($ a \equiv b \land b \equiv c \implies a \equiv c$)~\cite{WR83}. Grouping of points based on $\equiv$ forms equivalence classes. Similarity in network analysis occurs when two nodes fall in the same equivalence class. There are two fundamental approaches for constructing measures of network similarity: structural equivalence~\cite{LW71} and regular equivalence~\cite{newman2010networks}. There exists a hierarchy of these two equivalence concepts. Any structural equivalence is also regular equivalence, but not all regular equivalences are necessarily structural~\cite{hanneman2005introduction}.

Structurally equivalent nodes of a network must share the same neighbors. Formally is defined as follows~\cite{WR83}: if $G = (P, R)$ and $\equiv$ is an equivalence relation on $P$, then $\equiv$ is structural equivalence if and only if for all $a, b, c \in P$ such that $a \neq c \neq b$, $a \equiv b$ implies: (i) $aRb \leftrightarrow bRa$, (ii) $aRc \leftrightarrow bRc$; and (iii) $cRa \leftrightarrow cRb$. The undirected graph in~\figref{structural_eq} visually represents the concept, where nodes colored in red are the ones that are structurally equivalent.

Nodes that are regularly equivalent do not necessarily share neighbors, but have neighbors who are themselves similar. Formally is defined as follows~\cite{WR83}: if $G = (P, R)$ and $\equiv$ is an equivalence relation on $P$ then $\equiv$ is a regular equivalence if and only if for all $a, b, c \in P$, $a \equiv b$ implies: (i) $aRc \implies \exists d \in P, bRd \wedge d \equiv c$; and (ii) $cRa \implies \exists d \in P, dRb \wedge d \equiv c$. This concept is visually depicted on the same undirected graph in~\figref{regular_eq}, where regularly equivalent nodes are colored with the same color.

\subsection*{Measuring structural equivalence of nodes}
To find structurally equivalent nodes, one must compare each node with the rest and check which pairs of nodes have matching neighborhoods. This takes $\mathcal{O}(n^2)$ steps, as one is doing a pairwise comparison of $n$ nodes in the network. Most of the comparisons are needless as not every node is connected to everyone else. We should only compare neighborhoods of connected nodes, and that is achieved if we traverse through the edges of a network, which takes $\mathcal{O}(m)$ steps, where $m$ is the number of edges in the network. We consider edges $(x,y)$ one at a time and compare neighborhoods of nodes $x$ and $y$. We do not follow actual definition of structural equivalence, which gives binary answer (if neighborhoods match or not), but instead we measure to what extent they do match. The basic similarity measure just checks the size of the intersection of the two neighborhoods, i.e., the number of the common neighbors of two nodes~\cite{LW71}. The problem with this measure is that it is advantageously for nodes with bigger neighborhoods, which would therefore also have more common neighbors with the others. The example in~\figref{structural_eq_basic_measure_prob} shows the problem. By applying measure to pairs $(b,a)$ and $(c,a)$, the measure would return two and one common neighbor, respectively, and would mean that node $b$ is more structurally equivalent to node $a$ than node $c$ to node $a$. This is not true, as node $b$ has additional neighbor, uncommon to node $a$. Problem can be solved by using the Salton cosine~\cite{SMG83} similarity, which normalizes results with square root of degrees of connected nodes. If $A$ is adjacency matrix, where $A_{ij}$ indicates the number of neighbors between node $i$ and $j$, and $\Gamma_i$ is a set of neighbors of node $i$, and $k_i$ is the degree of node $i$, then cosine similarity $\sigma_{ij}$ is defined as:

\begin{equation*}
\sigma_{ij} = \cos \theta_{ij} = \frac{\sum_k A_{ik} A_{kj}}{  \sqrt{\sum_k \smash[b]{A^2_{ik}}} \sqrt{\sum_k \smash[b]{A^2_{jk}}} } =  \frac{| \Gamma_i \Gamma_j |}{ \sqrt{k_i k_j} }
\end{equation*}

Before measuring structural equivalence of nodes in the network \textit{npm}, we added additional edges. In a network of package dependencies such as \textit{npm}, a package depends not only on those packages that it points to but also on those packages that its dependencies point to. When you install a package from the \textit{npm}, the installer first installs dependencies of chosen package and then the package itself. But before dependencies are installed, it has to install their dependencies, and so on. So when you install a package, you get all its descendants. In that sense, a package depends on all descendants, therefore we added additional directed edges between a package and all its descendants. After-that we proceeded with measuring structural equivalence of nodes. For two nodes to even be considered structurally equivalent they have to share at least one common neighbor. In our case two packages must depend to at least  one same package. This is achieved if we pick each node and measure structural equivalence between pairs of its predecessors. This way a pair of predecessors will have at least picked node in common. When measuring structural equivalence of two nodes we first check if their neighbors overlap. If they do, then we mark those nodes as identical, otherwise we measure their cosine similarity. Afterwards  we remove any edges that connects examined pairs, because they no longer need to be compared. We keep picking nodes and measuring similarity of its predecessors until we visited all nodes and all edges are removed. From all measured pairs we build a new undirected weighted network, where  a pair of connected nodes is one of a measured pair, and each edge has a weight that is equal to a cosine similarity of connected nodes. Subsequently we contract nodes that we found to be identical, i.e., have identical neighbors. From the new network we extract similarity matrix, which we then give to a clustering method to find clusters of structurally equivalent nodes.

\subsection*{Measuring structural equivalence of nodes}

For finding regularly equivalent nodes, we used the implementation of CATREGE~\cite{BE93} algorithm in R programming language~\cite{sna}. It takes a graph and it assess how much are nodes regularly equivalent. For start, all nodes have the same role, therefore all are regularly equivalent. Then those nodes within the same role but different combination of neighbor types are re-allocated to different roles. Initially neighbor type is the pattern of in- and out-connections. This procedure is then iterated until all nodes within each role have same combination of neighbor types. The distance between nodes in this case is the inverse of the number of iterative refinements of the initial role required to allocate the nodes to regularly equivalent roles. The distance of 0 indicates nodes which belong to the same role. The algorithm gives results in a form of similarity/dissimilarity matrix, which can then be used with a clustering method. The algorithm has a cubic complexity, therefore it is applicable only on a smaller graphs that consist of no more than thousand nodes. 

\subsection*{Clustering with a distance matrix}
 We used k-medoids clustering algorithm~\cite{kaufman1987clustering}, which is similar to k-means algorithm. Both algorithms are partitioning technique of clustering that clusters the data set of $n$ objects into $k$ clusters known a priori. They can take matrix of distances between points and partition points into clusters by minimizing the distance between a point in a cluster and a point designated as the center of that cluster (medoid). In contrast to the k-means algorithm, k-medoids works with an arbitrary distance between points instead of just euclidean distance. The most common realization of k-medoid clustering is the Partitioning Around Medoids (PAM). For start, PAM randomly selects $k$ points as medoids. Then it associates each point to closest medoid and starts swapping medoids with non-medoids until cost of configuration stops decreasing. We used implementation of PAM in R programming language~\cite{pam}.
 
We applied PAM clustering on the similarity matrix of structural equivalence and dissimilarity matrix of regular equivalence. Beforehand, we transformed the similarity matrix of structural equivalence to dissimilarity matrix by subtracting 1 with cosine similarities. In this way, we got a cosine dissimilarity matrix. The matrices of both equivalence contain pairwise dissimilarity of all nodes in the network, therefore they take $O(n^2)$ space, where $n$ is the number of nodes. Consequently, we cannot work with the whole \textit{npm} network due to its big size. Therefore, we were obliged to sample the network and work with samples. We used simple random walk algorithm~\cite{	leskovec2006sampling} for sampling. We do not know how many groups of structurally equivalent nodes are there, therefore we need to run PAM algorithm multiple times and set different $k$ each time. Because of this time constrain, we sampled 1000 nodes. For finding groups of regularly equivalent nodes we sampled only 500 nodes due to cubic time complexity of CATREGE algorithm.

For determining $k$ we used the silhouette method~\cite{rousseeuw1987silhouettes}. For each node $i$, let $a(i)$ be the average dissimilarity of $i$ with all other nodes within the same cluster, and let $b(i)$ be the lowest average dissimilarity of $i$ to any other cluster, of which $i$ is not a member. The smaller the value $a(i)$, the better is a node assigned to the cluster. The cluster with this lowest average dissimilarity $b(i)$ is the next best cluster for the point $i$. Silhouette is defined as follows:

\begin{equation*}
s(i) = \frac{b(i) - a(i)}{max\{a(i),b(i)\}} = 
\begin{cases}
    1 - \frac{a(i)}{b(i)},& \text{if } a(i) < b(i) \\
    0,& \text{if } a(i) = b(i) \\
    \frac{b(i)}{a(i)} - 1,& \text{if } a(i) > b(i) \\
\end{cases}
\end{equation*}

We are interested when $a(i)$ is small, which means node $i$ is well matched to its cluster, and when $b(i)$ is bigger, which means node $i$ is badly fit to neighboring cluster. Thus $s(i)$ will be close to one. If it equals to zero, then node fits to two clusters equally well, and if it is negative, then a node is more fit to neighboring cluster. The average $s(i)$ over all nodes of a cluster is a measure of how tightly grouped all the nodes in the cluster are. Thus it is appropriately measure of how the nodes has been clustered and can used to determine the number of clusters within a network.

\section{Results}
\label{sec_res}

\subsection*{Regular equivalence classes}
\begin{figure}
\centering
\includegraphics[width=1.0\linewidth]{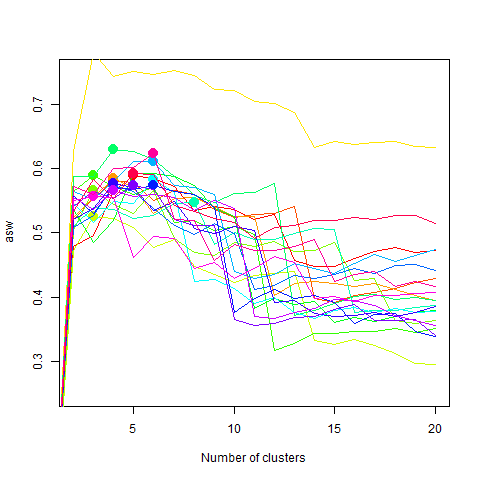}
\caption{Determining the number of regular equivalence classes (clusters) in the \textit{npm} network  with average silhouette value.}
\label{fig_re-avg-silhouette}
\end{figure}

\begin{figure*}
 \centering
 \begin{subfigure}{.3\textwidth}
   \centering
   \includegraphics[width=1.0\linewidth]{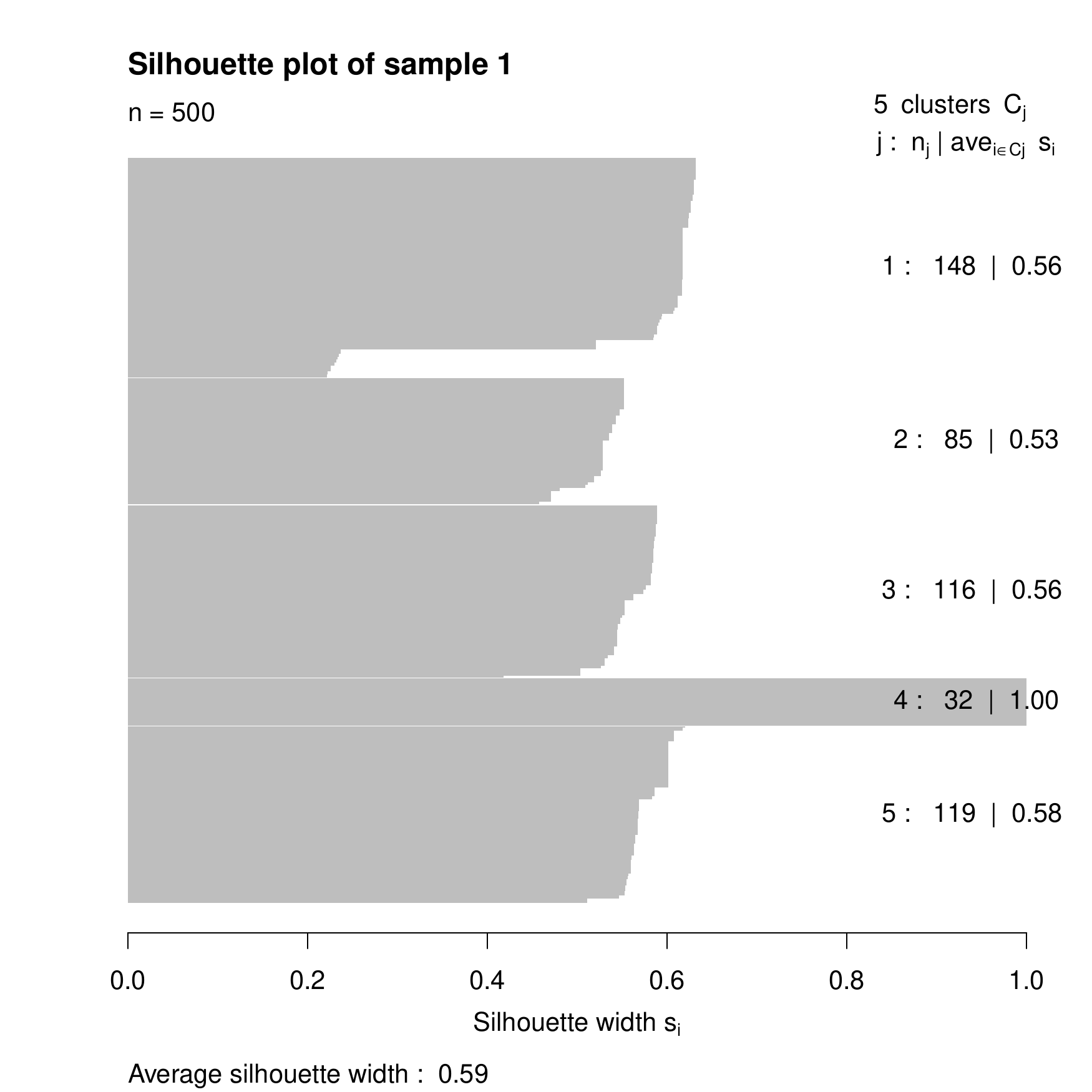}
 \end{subfigure}%
 \begin{subfigure}{.3\textwidth}
   \centering
   \includegraphics[width=1.0\linewidth]{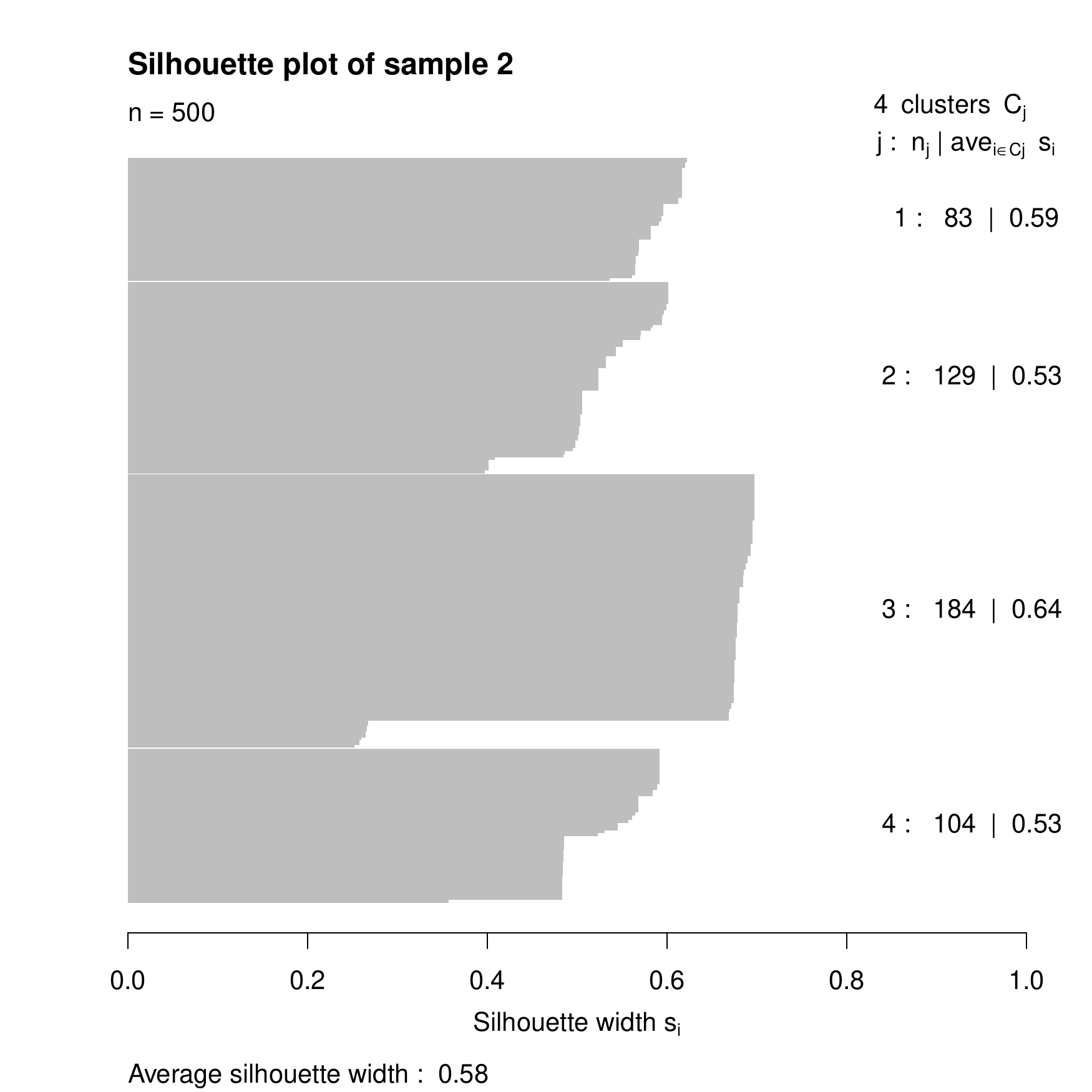}
 \end{subfigure}%
 \begin{subfigure}{.3\textwidth}
   \centering
   \includegraphics[width=1.0\linewidth]{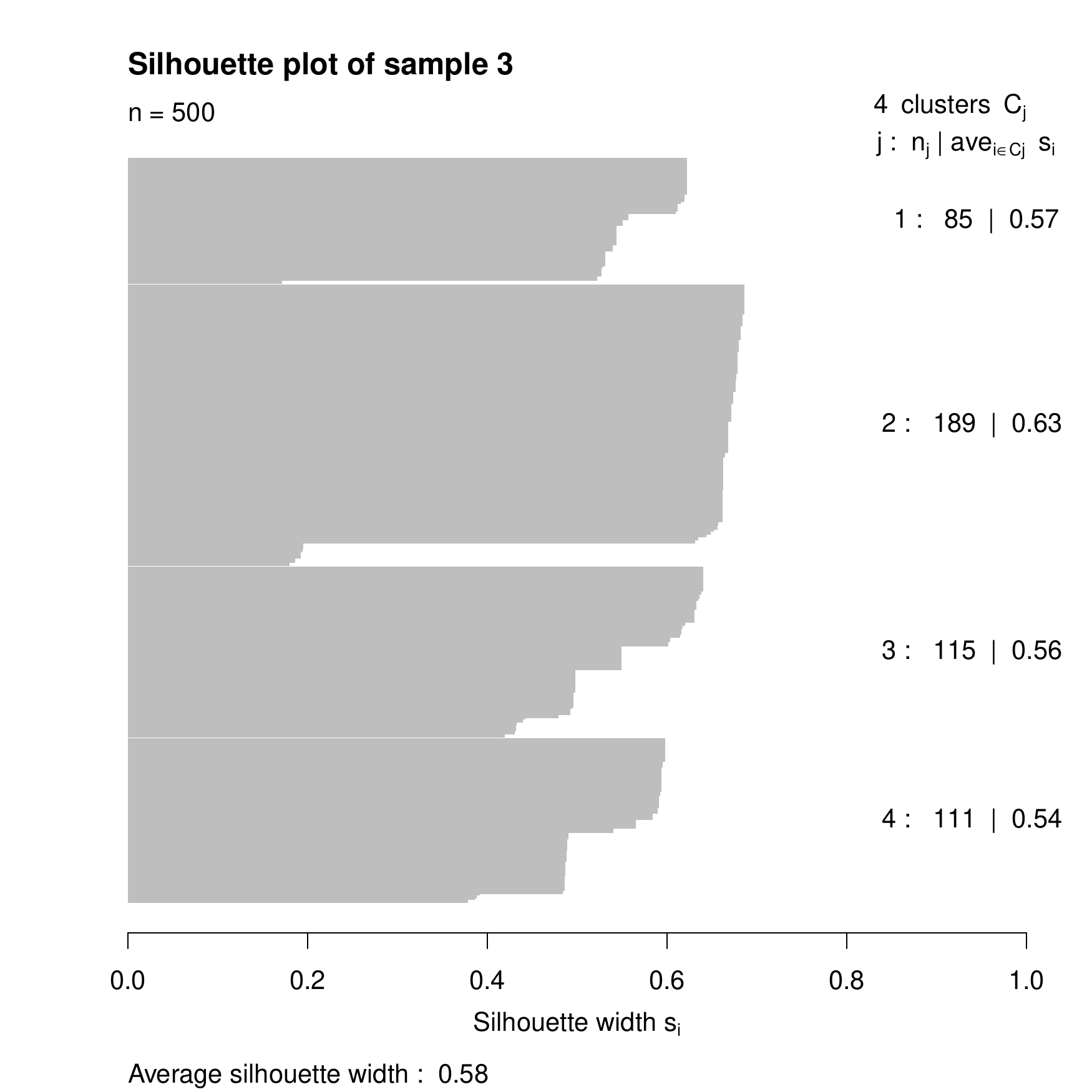}
 \end{subfigure}
 \caption{Silhouette plots of the first 3 samples for determining the number of regular equivalence classes in \textit{npm} network.}
 \label{fig_re-silhplots}
 \end{figure*}

\begin{figure*}
\centering
\begin{subfigure}{.3\textwidth}
  \centering
  \includegraphics[width=0.95\linewidth]{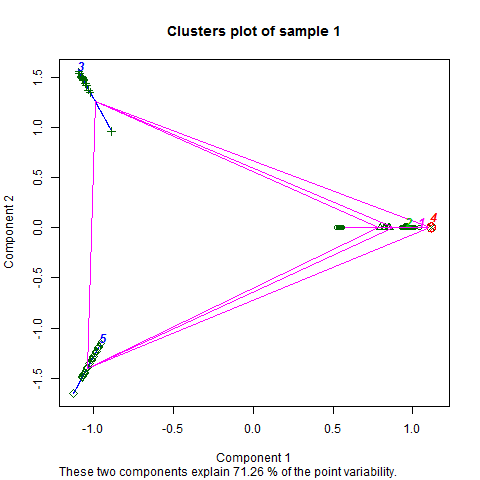}
\end{subfigure}%
\begin{subfigure}{.3\textwidth}
  \centering
  \includegraphics[width=0.95\linewidth]{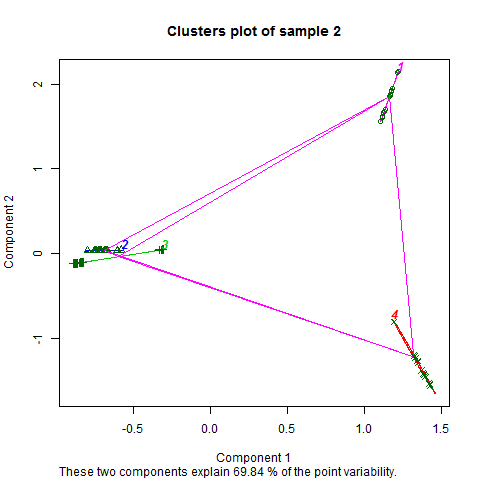}
\end{subfigure}%
\begin{subfigure}{.3\textwidth}
  \centering
  \includegraphics[width=0.95\linewidth]{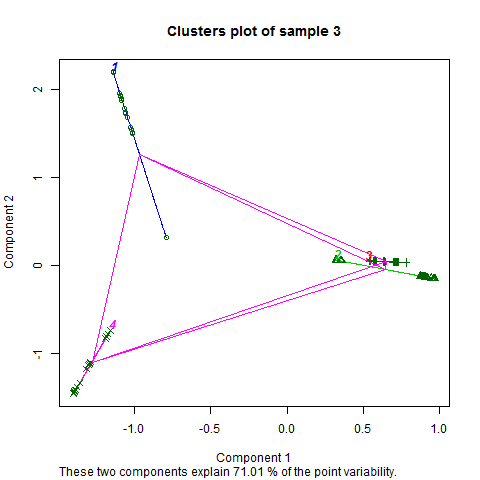}
\end{subfigure}
\caption{Clusters plots of the first 3 samples for determining the number of regular equivalence classes in \textit{npm} network.}
\label{fig_re-clusplots}
\end{figure*}

\begin{figure*}
\centering
\begin{subfigure}{.3\textwidth}
  \centering
  \includegraphics[width=0.95\linewidth]{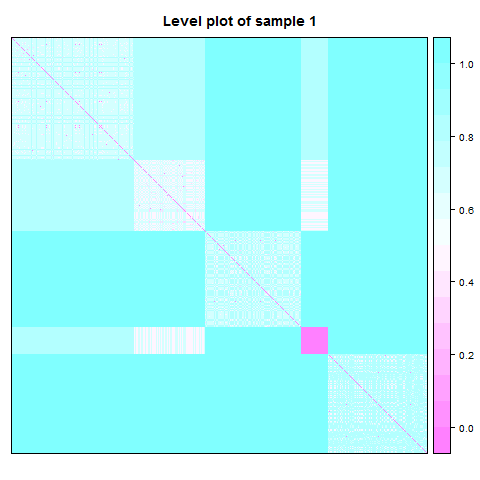}
\end{subfigure}%
\begin{subfigure}{.3\textwidth}
  \centering
  \includegraphics[width=0.95\linewidth]{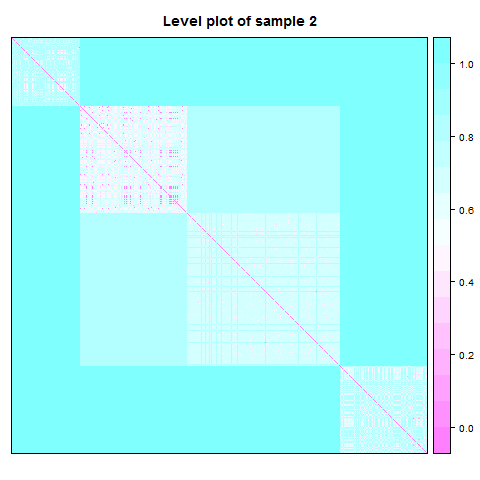}
\end{subfigure}%
\begin{subfigure}{.3\textwidth}
  \centering
  \includegraphics[width=0.95\linewidth]{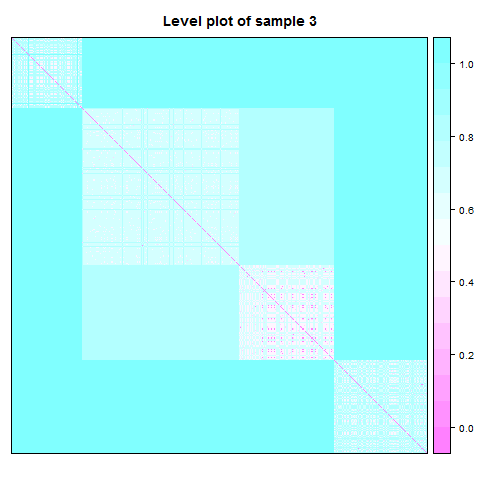}
\end{subfigure}
\caption{Level plots (block modeling) of the first 3 samples for determining the number of regular equivalence classes in \textit{npm} network.}
\label{fig_re-levelplots}
\end{figure*}

For regular equivalence we sampled the \textit{npm} network 20 times by randomly visiting 500 nodes. Within samples the mean number of edges is 1350 (std. dev. is 158). Each sample was clustered with PAM and $k \in [2,20]$. The average silhouette value for each sample and each $k$ is shown in~\figref{re-avg-silhouette}. The optimal number of clusters is $~\approx$ 4, which corresponds to the number of regular equivalence classes. The average silhouette value over all nodes is $~\approx 0.6$, therefore nodes fit quite well to assigned clusters and regular classes. The silhouette plots of the first 3 samples is shown in~\figref{re-silhplots}, where we can see that average silhouette value of nodes within the same cluster is more than 0.5, which is the another sign of the good clustering. Another kind of plots, called cluster plots, of the first 3 samples is shown in~\figref{re-clusplots}. These plots show clustered nodes in the first 2 principal components, which capture$~\approx$ 80\% of variance of dissimilarity matrix. We can see that nodes can actually be clustered in 3 clusters, and thus 2 or more clusters can be regarded as the same regular equivalence class. This is also nicely shown in level plots (block modeling) in~\figref{re-levelplots}, where there are 3 prominent blocks which represent 3 regular equivalence classes. After examining the nodes in clusters, we found out that a cluster have either nodes that have only in-connections, or out-connection, or both in- and out-connections. There is a single cluster of nodes with only in-connections and a single cluster of nodes with only out-connections per sample. The rest clusters have nodes with both connections. The reason the nodes are not clustered in a single cluster is because they can be then further nicely divided by the scale of in- and/or out-connections. Our hypothesis states that there are following roles in the software networks: a group of core package with only in-connections, a group of popular packages with both connections, and a group of packages with only out-connections that supplement popular packages. A cluster of nodes with only in-connections represents a group of core packages, a cluster of nodes with both connections where there are more out- than in-connections represents a group of popular packages, and the rest clusters represent supplementary packages.

\subsection*{Structural equivalence classes}
\begin{figure}
\centering
\includegraphics[width=1.0\linewidth]{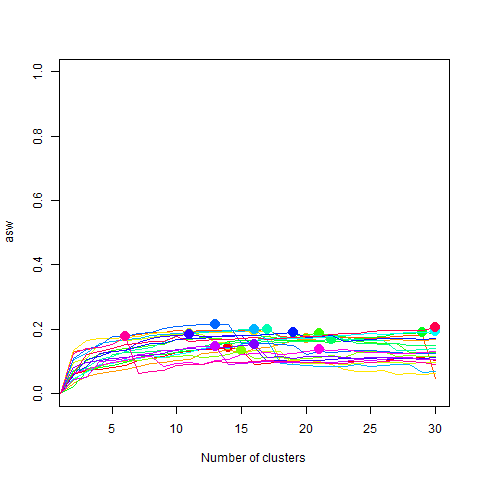}
\caption{Determining the number of structural equivalence classes (clusters) in the \textit{npm} network with average silhouette value.}
\label{fig_se-avg-silhouette}
\end{figure}

\begin{figure*}
\centering
\begin{subfigure}{.3\textwidth}
  \centering
  \includegraphics[width=0.95\linewidth]{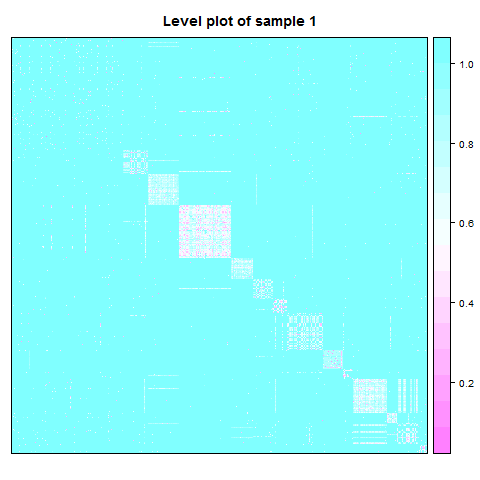}
\end{subfigure}%
\begin{subfigure}{.3\textwidth}
  \centering
  \includegraphics[width=0.95\linewidth]{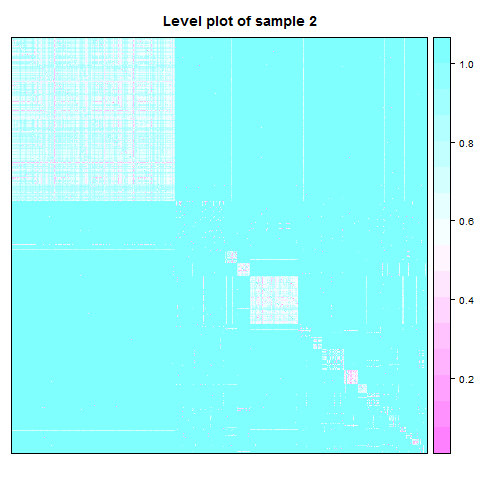}
\end{subfigure}%
\begin{subfigure}{.3\textwidth}
  \centering
  \includegraphics[width=0.95\linewidth]{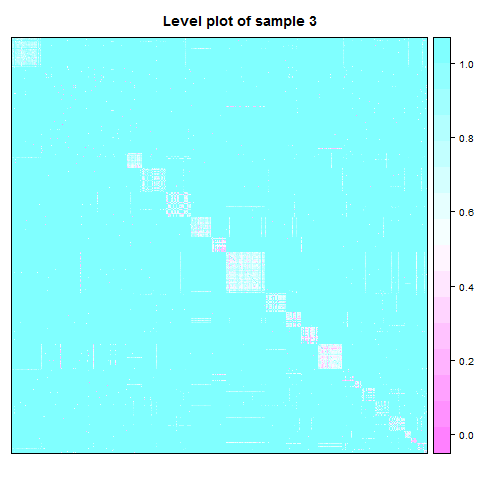}
\end{subfigure}
\caption{Level plots (block modeling) of the first 3 samples for determining the number of structural equivalence classes in \textit{npm} network.}
\label{fig_se-levelplots}
\end{figure*}

For structural equivalence we sampled the \textit{npm} network 20 times by randomly visiting 1000 nodes. Within samples the mean number of edges is 8185 (std. dev. is 1711), the mean number of structurally equivalent node is 761 (std. dev. is 40), the mean number of edges between structurally equivalent nodes is 30772 (std. dev. is 13983), and the mean number of identical nodes is 270 (std. dev. is 40). Each sample was clustered with PAM and $k \in [1,30]$. The average silhouette value for each sample and each $k$ is shown in~\figref{se-avg-silhouette}. The optimal number of clusters cannot be determined because it differs from sample to sample, and therefore we cannot asses how many structural equivalence classes are there in the \textit{npm} network. Although the average silhouette value of the best clustering per sample is small, we can see nice grouping of nodes into blocks in the level plots in~\figref{se-levelplots}. These groups within the \textit{npm} network could theoretically correspond to different sub-ecosystems, i.e.,  different types of software.

\section{Conclusions}
Nodes of the \textit{npm} network were grouped by two equivalence relation: structural and regular equivalence. This was achieved by first measuring to what extent are pairs of nodes structurally/regularly similar/dissimilar. From pairwise similarities/dissimilarities we constructed dissimilarity matrix, which was then given to PAM clustering algorithm - the implementation of k-medoids. Because we had to work with matrix that takes quadratic space depending on a number of node in a network, we were obliged to work with a subgraphs. Due to this and other additional constrains, we could process at most thousand nodes per sample. Nonetheless, we successfully showed that regular equivalence relation when applied to the network indeed groups nodes into roles that we were seeking for. By applying structural equivalence to the network we found many different groups that could theoretically correspond to different types of software. 

\label{sec_conc}


\bibliographystyle{abbrv}
\bibliography{refs}

\end{document}